# Interplay between MRI-based axon diameter and myelination estimates in macaque and human brain

Ting Gong[a*], Chiara Maffei[a], Evan Dann[a], Hong-Hsi Lee[a], Hansol Lee[a], Jean C. Augustinack[a], Susie Y. Huang[a], Suzanne N. Haber[b,c], Anastasia Yendiki[a]

a. Martinos Center for Biomedical Imaging, Massachusetts General Hospital and Harvard Medical School, Charlestown, MA, United States
b. Department of Pharmacology and Physiology, University of Rochester, Rochester, NY, United States
c. McLean Hospital, Belmont, MA, United States

*Correspondence; Email: tgong1@mgh.harvard.edu

**Abstract**
Axon diameter and myelin thickness affect the conduction velocity of action potentials in the nervous system. Imaging them non-invasively with MRI-based methods is thus valuable for studying brain microstructure and function. Electron microscopy studies suggest that axon diameter and myelin thickness are closely related to each other. However, the relationship between MRI-based estimates of these microstructural measures, known to be relative indices, have not been investigated across the brain mainly due to methodological limitations. In recent years, studies using ultra-high gradient strength diffusion MRI (dMRI) have demonstrated improved estimation of axon diameter index across white-matter (WM) tracts in the human brain, making such investigations feasible. In this study, we aim to investigate relationships between tissue microstructure properties across white-matter tracts, as estimated with MRI-based methods. We collected dMRI with ultra-high gradient strength and multi-echo spin-echo MRI on ex vivo macaque and human brain samples on a preclinical scanner. From these data, we estimated axon diameter index, intra-axonal signal fraction, myelin water fraction (MWF) and aggregate g-ratio and investigated their correlations. We found that the correlations between axon diameter index and other microstructural imaging parameters were weak but consistent across WM tracts in samples estimated with sufficient signal to noise ratio. In well-myelinated regions, tissue voxels with larger axon diameter indices were associated with lower packing density, lower MWF and a tendency of higher g-ratio. We also found that intra-axonal signal fractions and MWF were not consistently correlated when assessed in different samples. Overall, the findings suggest that MRI-based axon geometry and myelination measures can provide complementary information about fiber morphology, and the relationships between these measures agree with prior electron microscopy studies in smaller field of views. Combining these advanced measures to characterize tissue morphology may help differentiate tissue changes during disease processes such as demyelination versus axonal damage. The regional variations and relationships of microstructural measures in control samples as reported in this study may serve as a point of reference for investigating such tissue changes in disease.

# 1. Introduction

Axon diameter and myelin thickness are closely related features of tissue microstructure that affect the conduction velocities of neuronal signals (Waxman, 1980) and thus relate to brain function (Caminiti et al., 2013, 2009; Liewald et al., 2014; Tomasi et al., 2012). Precise measurement of such microstructural features requires nanometer resolution. With electron microscopic techniques, an almost linear relationship between axon diameter and myelin thickness has been established (Hildebrand and Hahn, 1978; Keyserlingk and Schramm, 1984). However, due to the challenges of tissue preparation and image acquisition, storage, and analysis, this type of study is limited to small fields of view, in the order of hundreds of micrometers (Foxley et al., 2021). Thus, investigating the relationship between axon diameter and myelin thickness across entire white-matter (WM) tracts remains out of reach with such methods.

Non-invasive, MRI-based techniques for probing white-matter microstructure offer the benefit of whole-brain coverage, and the potential to be used as *in vivo* biomarkers in basic and clinical neuroscience. Microstructural modeling based on MRI does not allow *absolute* measurements of axon diameter and myelin density, but it does provide *relative* "indices" that can be used to compare microstructure across brain regions or populations. The sensitivity of such measurements has been evolving rapidly with recent technical advancements. Studying how these advanced, MRI-based estimates of axonal diameter and myelination vary and relate to each other across WM tracts in the healthy brain is therefore important for better characterizing these MRI-based measures, and for interpreting the findings of studies that use these measures in healthy or diseased populations.

Axon diameter index is commonly estimated through microstructure modelling with diffusion weighted (DW) MRI. DW MRI is sensitive to the Brownian motion of water molecules in biological tissue, where the root mean squared displacement is at micrometer length scale at typical observation time, providing unique opportunities to estimate underlying tissue microstructure (Alexander et al., 2019; Novikov et al., 2019). Tissue microstructure is inferred by fitting the DW measurements to biophysical models representing tissue morphology. In representing signals from intra-axonal space, axons are modelled as cylinders with restricted water diffusion in the perpendicular direction (Assaf et al., 2004; Assaf and Basser, 2005; Neuman, 1974; Van Gelderen et al., 1994). This was first introduced to estimate the distribution of axon diameter indices through the gamma function approximation with the AxCaliber model in the spinal cord (Assaf et al., 2008) and corpus callosum (Barazany et al., 2009). This method requires the DW signals to be measured perpendicular to the fiber orientation at different diffusion times and gradient strengths, hence is only applicable to regions with known fiber orientation. The requirement of known fiber orientation is obviated by the ActiveAx method, with orientation invariant acquisition on several diffusion wave vector (q)-shells through an optimal experimental design (Alexander, 2008; Alexander et al., 2010; Zhang et al., 2011). This method estimates the major fiber orientation and the axon diameter index in the model and was later extended to consider dispersion around the fiber orientation (Alexander, 2008; Alexander et al., 2010; Zhang et al., 2011). Similar to ActiveAx, the TractCaliber method uses a q-shell acquisition from which the major fiber orientation is estimated. It then predicts the signals perpendicular to the fiber orientation and estimates the axon diameter index from the predicted signals (Huang et al., 2020). Improvements to the signal modelling in the extra-axonal space have also been considered (De Santis et al., 2016). These methods, however, assume a single fiber direction and are not applicable in regions with crossing fibers. As a result, applications of axon diameter indices have so far focused mostly on areas with coherent fibers, such as the midline of the corpus callosum (Genc et al., 2023; Huang et al., 2019).

Recent developments of axon diameter modelling have focused on using powder-averaged DW signals from the spherical mean technique (Kroenke et al., 2004) to factor out the effect of fiber orientations. These methods use ultra-high diffusion weighting (b-values) to increase sensitivity to small axons across whole-brain WM (Andersson et al., 2022; Fan et al., 2020; Pizzolato et al., 2023; Veraart et al., 2020; Warner et al., 2023). Such measurements have been made much more practical by human MRI scanners with ultra-strong gradients, allowing for shorter diffusion and echo times and higher signal to noise ratio (SNR) in the measurements (Huang et al., 2021, 2014; Jones et al., 2018; Mcnab et al., 2013). Axon diameter modelling often assumes that the axons are myelinated. However, a range of myelination exists in the brain and the relationships between these axon diameter indices and myelination across the brain remain largely unexplored.

Myelin water has very short T2 (~10 ms), making it invisible at typical echo times (>50 ms) in DW MRI (Mackay et al., 1994). Therefore, its quantification is often achieved through other MRI techniques (Mackay and Laule, 2016; Sled, 2017). A well-established method is myelin water imaging through T2 spectrum analysis (Whittall and MacKay, 1989), where the short-T2 signal component is attributed to myelin water. The myelin water fraction (MWF) can then be extracted as a measure reflecting the amount of myelin wrapped around axons. Currently, spatial variations of myelin content have been reported across the brain (Cercignani et al., 2017; Dean et al., 2016) and the mechanism behind population differences in myelination is an active area of research in neurological conditions, brain development, maturation, and aging (Baum et al., 2022; Cábez et al., 2023; Call and Bergles, 2021; Clark et al., 2021; Grotheer et al., 2022).

Combining axonal and myelination measures is critical for fully characterizing morphological and functional properties of tissue. For example, the fiber g-ratio, defined as the ratio between the inner axon diameter and outer fiber diameter (including myelin sheath), is shown to determine the conduction velocity in histological studies (Castelfranco and Hartline, 2016; Ritchie, 1982; Sanders and Whitteridget, 1946; Waxman, 1980). For MRI-based methods, the tissue aggregate g-ratio has been calculated from myelin volume fraction (MVF) and axon volume fraction (AVF) (Campbell et al., 2018; Mohammadi and Callaghan, 2021; Stikov et al., 2015), which are typically approximated by myelin-sensitive measures and intra-axonal signal fractions estimated from DW MRI (Berg et al., 2022), such as NODDI (Zhang et al., 2012). The MR-based g-ratio has been investigated in early brain development (Dean et al., 2016), healthy aging (Bouhrara et al., 2021; Cercignani et al., 2017) and the spinal cord (Duval et al., 2016). However, the direct relationship between axon diameter index and myelination and how they associate with g-ratio has not been investigated across WM tracts.

As the methodology for estimating axon diameter index has evolved in recent years, revisiting the relationships of axonal and myelination measures across WM tracts is of importance. In this study, we estimate axon diameter index, intra-axonal signal fraction, MWF and g-ratio with data collected on ex vivo macaque brains and a human brain tissue block with a pre-clinical MRI scanner. In estimating axon diameter index, we introduce a multicompartment model with DW MRI data acquired at multiple b-shells and ultra-high b-values that is tailored to ex vivo tissue. We further validate parameter estimation with synthesized data. Using ex vivo data collected with different signal-to-noise ratio (SNR) levels, we assessed the regional variations of these microstructure measures across WM tracts and samples and investigate the correlations between them. The relationships reported in this healthy tissue study can serve as a baseline for investigations of changes in these measures in various applications. For example, some applications involve coexisting processes that cause tissue changes, such as axonal damage and demyelination in multiple sclerosis (Hill et al., 2021; Huang et al., 2019; Nedjati-Gilani et al., 2017) or axonal growth and myelination in brain development and maturation (Genc et al., 2023). We anticipate that combining axonal geometry and myelination measures will help improve characterization of tissue changes in these applications, facilitated by ultra-high-gradient MRI scanners for in vivo imaging (Huang et al., 2021).

## 2. Materials & Methods
### 2.1 Sample preparation
#### 2.1.1 Macaque brains
Four adult male macaques (Macaca fascicularis) were anesthetized and perfused with saline followed by a 4% paraformaldehyde/1.5% sucrose solution in 0.1 m PB, pH 7.4. Brains were postfixed overnight and cryoprotected in increasing gradients of sucrose (10%, 20%, and 30%). Surgery and tissue preparation of macaque brain samples were performed at the University of Rochester and details of the procedures are described in previous studies (Grisot et al., 2021; Safadi et al., 2018).

#### 2.1.2 Human tissue block
One tissue block was extracted from the left-brain hemisphere of a 60-year-old female (cause of death: adenocarcinoma of pancreas; postmortem interval: 2 hours) (Williams et al., 2023). The hemisphere was fixed in 10% formalin for at least two months. A 7.5x3.5x1.5 cm tissue block was extracted from a coronal slab in the frontal lobe, containing segments of the anterior limb of the internal capsule and anterior segments of the superior longitudinal fasciculus.

### 2.2 Data acquisition
We acquired data for the whole macaque brains and human tissue block using a small-bore 4.7 T Bruker BioSpin MRI system equipped with maximum gradient strength of 660 mT/m. The tissue samples were packed in Fomblin (Solvay, Italy) to eliminate background signal. Two of the macaque brains were scanned with a different protocol, featuring higher SNR, to investigate the effect of SNR on microstructural parameter estimation. A summary of the data acquisition parameters that differed across samples, along the SNR and T2 values measured from each dataset, is given in Table 1. Details about estimation of SNR and T2 is included in supplementary materials S1.

#### 2.2.1 Diffusion MRI
The DW images were collected using a 3D echo-planar imaging sequence at 0.5 mm isotropic resolution. The TR for all scans was 500 ms, with TE > 50 ms for samples acquired with two segments and two averages and TE = 33 ms for samples acquired with eight segments (see Table 1). The FOV and image size were adjusted for each sample depending on the sample size; about 6/8 partial Fourier was applied to the phase encoding direction. DW images were collected with 8 different b-shells at 1, 2.5, 5, 7.5, 11.1, 18.1, 25 and 43 ms/$\mu m^2$, where 12 (for b<=7.5 ms/$\mu m^2$) or 32 (for b>=11.1 ms/$\mu m^2$) gradient directions were sampled over the sphere; one b=0 image was acquired before each b-shell. The separation between diffusion gradient pulses was Δ=15 ms and the duration of diffusion gradient pulses was δ=11 ms. The total acquisition time was 14 hours for a macaque brain and 16 hours for the human tissue block.

For the human tissue block, the whole left hemisphere had previously been scanned on a 3.0 T MRI scanner (0.75 mm isotropic; 12 images at b = 0 and 90 gradient directions at b = 3.8 ms/$\mu m^2$). We used this whole-hemisphere scan here only to perform tractography and locate tracts of interest in the tissue block, as described in a later section.

### 2.2.2 Myelin water imaging

We collected multi-slice multi-echo (MSME) images using the Carr Purcell Meiboom Gill (CPMG) sequence with slice selective RF pulses to quantify myelin water. The images were acquired with 3D encoding and 0.5 mm isotropic resolution without partial Fourier or signal averaging. For macaque samples, the spin-echo images were collected with a TR of 3000 ms, either at 20 echo times with an equal echo spacing of 8 ms, or at 32 echo times with echo spacing 6 ms (see Table 1). Total imaging time was about 10 hours. For human samples, the MSME images were collected at 40 echo times from 5-200ms with an equal spacing of 5ms and TR of 2000 ms. Total imaging time was 5 hours.

Table 1. Summary of acquisition protocol, SNR of b=0 images from the DW scan, and T2 relaxation constant calculated from MSME data in each sample. (We note that direction averaging from N measurements with the same b-value would lead to an SNR improvement by a factor of sqrt(N) for spherical mean measurements over the SNR level on DW images. This factor ranges from 2.8 for b=0, where N=8, to 5.7 for high b-values, where N=32.)

|  |  | **Macaque 1** | **Macaque 2** | **Macaque 3** | **Macaque 4** | **Human block** |
|---|---|---|---|---|---|---|
| **dMRI scan** | TE | 52 ms | 52 ms | 33 ms | 33 ms | 55ms |
|  | Number of segments | 2 segments, 2 averages | 2 segments, 2 averages | 8 segments | 8 segments | 2 segments, 2 averages |
|  | Mean WM SNR | 23.8 | 31.4 | 54.1 | 63.3 | 37.7 |
| **MSME scan** | Number of echoes | 20 | 20 | 32 | 32 | 40 |
|  | TEs | 8: 8 :160 ms | 8: 8: 160 ms | 6: 6: 192 ms | 6: 6: 192 ms | 5: 5: 200 ms |
|  | Mean WM T2 | 31.2 ms | 37.1 ms | 34.3 ms | 33.8 ms | 71.6 ms |

### 2.3 Data preprocessing

We first denoised the DW MRI and MSME data, as noise can introduce bias to quantitative parametric mapping. In typically reconstructed and saved magnitude images, the Gaussian noise in the real and imaginary channel introduces Rician bias to the magnitude signals, which is more prominent for low-SNR data, such as in high b-value DWIs and T2-weighted images with longer echo times in the MSME scan. A previous study has suggested that Rician noise can introduce bias to axon diameter index estimation, and extracting real-valued dMRI data with only Gaussian additive noise presented can improve the accuracy of axon diameter index estimation (Fan et al., 2020) and intra-neurite fraction estimation (Ianuş et al., 2022). Denoising images can further improve the SNR, hence the precision of parameter estimation. Other studies have shown that denoising complex images is preferable to magnitude images in both dMRI and myelin water imaging (Does et al., 2019; Manzano Patron et al., 2024).

We evaluated two strategies to reduce bias and noise in the datasets in addition to applying Marchenko-Pastur (MP)-PCA denoising to magnitude images (Veraart et al., 2016b, 2016a). 1. Combining the real and imaginary channel of the complex images for Gaussian noise level estimation; denoising both channels and combining them to obtain magnitude images. 2. Obtaining real-valued images from the complex data as in the background phase filtering procedure introduced in (Eichner et al., 2015; Fan et al., 2020); denoising the real-valued images and discarding the imaginary channel containing mostly background noise. These two denoising strategies require complex MRI datasets.

The denoised DW datasets were checked for signal drift (Vos et al., 2017) (supplementary S3), followed by eddy current correction (Andersson et al., 2003), gradient orientation correction (Jeurissen et al., 2014), and bias field correction (Tustison et al., 2010). We obtained powder-averaged signals for axon diameter index estimation (Andersson et al., 2022; Fan et al., 2020; Veraart et al., 2020). This was done by averaging the DW signals across all gradient directions at each b-shell. We then normalized the signals by the mean b=0 signal prior to axon diameter index estimation.

### 2.4 Microstructural modelling
#### 2.4.1 Tissue model for estimating axon diameter index
We fitted the powder-averaged DW signal at each WM voxel to a three-compartment tissue model, including intra-axonal and extra-cellular compartments, as well as a dot compartment relevant to ex vivo imaging (Alexander et al., 2010; Panagiotaki et al., 2012). The dot compartment represents isotropically restricted water molecules, from which the signals do not decay even at very high b-values. The biological origin of the dot compartment remains poorly understood. A few studies have discussed potential contributing factors, including water restriction in glial cells (Stanisz et al., 1997) and cell nuclei and vacuoles (Andersson et al., 2020). This compartment is often substantial in fixed ex vivo tissue, suggesting that tissue changes due to death and fixation may further contribute to it. Although signal contribution from the dot compartment tends to be negligible in typical in vivo data, it has also been demonstrated in vivo using ultra-strong gradients and special diffusion encoding, especially in the cerebellar gray matter (Tax et al., 2020). In our ex vivo data, the signal decayed to a non-zero value for high b-values, hence including the dot compartment was necessary to obtain a good fit of the multi-compartment model. We initially included a compartment for free water but, after determining that no such compartment was present in our ex vivo samples, we omitted it from all subsequent analyses.

The three tissue compartments included in our model contribute to the powder-averaged signal decay (normalized to the b=0 signal) as a function of b-value as follows:

$$S(b) = f_{ia}S_{ia}(b; D_{\parallel}^{ia}, D_0, d_a) + f_{ec}S_{ec}(b; D_{\perp}^{ec}, D_{\parallel}^{ec}) + f_{dot} \quad [1],$$

where the b-value is determined by the diffusion gradient pulse width $\delta$, separation $\Delta$, strength $G$ and gyromagnetic ratio $\gamma$ as $b = (\gamma\delta G)^2(\Delta - \delta/3)$ in a pulsed gradient spin echo sequence; $f_{ia}$, $f_{ec}$ and $f_{dot}$ are the signal fractions of intra-axonal, extra-cellular and dot compartments, with $f_{ia} + f_{ec} + f_{dot} = 1$; and $S_{ia}$ and $S_{ec}$ are the signal decay functions for the intra-axonal and extra-cellular compartments (their details can be found in Appendix I). Briefly, signal decay of the intra-axonal space (IAS) is modelled by assuming impermeable cylinders with a Gaussian phase distribution approximation perpendicular to the fiber (Van Gelderen et al., 1994). The simple cylinder model assumes no changes of trajectory and curvature along axons on the length scale of the measured diffusion, thus the intra-axonal parallel diffusivity $D_\parallel^{ia}$ is equal to the intrinsic diffusivity $D_0$; the model is parameterized by tissue parameters of $D_\parallel^{ia}$ and axon diameter index $d_a$. The IAS model is detailed in (Andersson et al., 2022; Fan et al., 2020), and has been demonstrated to achieve axon diameter index estimation at highly aligned fiber and crossing fiber regions in simulations of realistic fibers segmented from mouse brain (Lee et al., 2020a), vervet monkey brain (Andersson et al., 2022), and human brain (Lee et al., 2024), and in vivo MRI of human brain (Fan et al., 2020). Signal decay from the extra-cellular compartment is modelled as an anisotropic Gaussian parameterized by parallel ($D_\parallel^{ec}$) and perpendicular diffusivity ($D_\perp^{ec}$).

### 2.4.2 Model fitting

Previous studies have adopted some simplifications, reducing the number of parameters to estimate tissue parameters of interest (i.e., $d_a$ and $f_{ia}$) robustly. Common simplifications include assumptions that the parallel diffusivities (intra-axonal and extra-cellular) are equal to the intrinsic diffusivity, which can be fixed to typical values measured in ex vivo or in vivo tissue. Here, we assume $D_\parallel^{ec} = D_\parallel^{ia} = D_0$. However, we estimated the parallel diffusivity from the tissue model. Hence the tissue parameters we fitted were $\theta = (f_{ia}, d_a, D_\parallel^{ia}, D_\perp^{ec}, f_{dot})$. We constrained $D_\perp^{ec}$ to be smaller than $D_\parallel^{ec}$ by representing $D_\perp^{ec}$ as $D_\parallel^{ec}$ multiplied by a fraction between [0, 1]. This models the assumption that water diffusion in the extra-cellular space of white matter tissue is hindered and anisotropic (Jespersen et al., 2010), hence $D_\perp^{ec}$ is smaller than $D_\parallel^{ec}$.

We used the Markov Chain Monte Carlo (MCMC) method to sample the posterior distribution of model parameters. We used uniform priors for diameter $d_a$ ~ [0.1-10] μm, parallel diffusivity $D_\parallel^{ia}$ ~ [0.01, 0.9] μm²/ms, the fraction of extra-cellular perpendicular diffusivity to parallel diffusivity ~ [0, 1] and the noise level $\sigma$ ~ [0.001-0.1]. Priors for the signal fractions ($f_{ia}\ and\ f_{dot}$) were in the range [0, 1], and their sum was constrained in [0, 1], effectively following a Dirichlet distribution. Proposal distributions were Gaussians with small standard deviations. We assumed Gaussian noise, hence the likelihood of measuring powder-averaged DW signal $\tilde{S}$ under modelled signal $S$ at noise level $\sigma$ is:

$$L(\tilde{S}|S, \sigma) = (2\pi\sigma^2)^{-\frac{n}{2}} \exp\left(-\sum_{i=1}^{n} \frac{(\tilde{S}(i) - S(i))^2}{2\sigma^2}\right)\ [2],$$

where $S$ is the prediction from Eq. [1] given acquisition parameters and estimated tissue parameters; $n$ is the number of DW measurements equivalent to the number of non-zero b-values. The log-likelihood ratio was updated using a Metropolis-Hastings sampling algorithm. We used an initial burn-in period of 20,000 iterations and a sampling interval of 100 to gather 500 samples in each run. The estimated tissue parameters were calculated as the mean values of the samples.

When compared to fixing the diffusivity $D_\parallel^{ia}$ to typical values (0.6 $\mu m^2$ /ms for ex vivo tissue and 1.7 $\mu m^2$ /ms for in vivo tissue), we found that fitting all parameters improves the quality of fit by inspecting the Bayes factor. We also found, however, that this introduces a higher uncertainty of estimated $f_{ia}$ by inspecting the standard deviations of posterior distribution. We thus repeated the MCMC process twice by first sampling the distributions of all parameters and then, in the second run, fixing $D_\parallel^{ia}$ and $D_\perp^{ec}$ to their posterior means voxel by voxel and sampling only the distributions of $f_{ia}, d_a$, and $f_{dot}$. The second MCMC run gives roughly the same likelihood of measurements and lower parameter uncertainty for $f_{ia}$ and $d_a$, compared to the first run.

### 2.4.3 MWF estimation

For myelin water imaging data, we estimated the T2 spectrum from the denoised data using the Julia toolbox for Decomposition and Component Analysis of Exponential Signals (DECAES) (Doucette et al., 2020). We used a regularized non-negative least squares method (Hansen, 2006) with calibration to imperfect refocusing pulses due to B1 field inhomogeneity (Prasloski et al., 2012). The spectrum was defined at 100 T2 values logarithmically spaced between 4 ms and 200 ms for the macaque samples, and between 6ms and 250 ms for the human sample. From the T2 spectrum, MWF was calculated as the signal fraction from 8ms < T2 < 15 ms for the macaque brains and 5ms < T2 < 30 ms for the human tissue block, after inspecting the spectrum. Details on how we determined the range of T2 for myelin water can be found in Supplementary materials S2.

### 2.5 Cross-modal registration

We registered the parameter maps estimated from dMRI into the space of the MSME scan of the same sample, so that axon diameter index, intra-axonal signal fraction and MWF maps were aligned for further analysis. We performed this registration with the advanced normalization tools (ANTs) (Avants et al., 2014) for both macaque brains and human tissue block. We chose one T2-weighted image from the MSME scan that has similar TE to the dMRI scan as the target for this registration. This is because the MSME images are less distorted than the dMRI scan, hence this step also serves as distortion correction for dMRI-derived maps. Before registration, the T2-weighted image from the chosen echo was masked to extract brain voxels and corrected for bias field to remove spatial intensity variation using the N4 bias correction in ANTs (Tustison et al., 2010). The mean b=0 image from dMRI was then non-linearly co-registered to the corrected T2-weighted image. The resulting warp field was applied to microstructure parameters estimated in dMRI space.

### 2.6 G-ratio calculation

After registration, we calculated the g-ratio from the intra-axonal water fraction estimated from the tissue model of diffusion and the MWF estimated from MSME data. The aggregate g-ratio is defined in previous studies (Stikov et al., 2015; West et al., 2015) as:

$$g = \sqrt{\frac{1}{1+MVF/AVF}} \quad [3],$$

where the myelin volume fraction (MVF) can be calculated from myelin sensitive measures and the axon volume fraction (AVF) can be calculated by combining MVF with DW MRI. In our study, we used

MWF to quantify myelin; following calibration factors for ex vivo tissue (West et al., 2018), MVF can be calculated from MWF estimates as:

$$MVF = \frac{MWF \times 0.859}{MWF \times 0.384 + 0.475} \quad [4]$$

The AVF was approximated by the absolute intra-axonal water fraction from the compartment model of diffusion as:

$$AVF = (1 - MVF) \times f_{ia} \quad [5]$$

Here $f_{ia}$ is the intra-axonal water fraction estimated from the tissue model with DW MRI.

### 2.7 Tracts and regions definition

For the human sample, we identified the main tract segments that pass through the human tissue block. Tractography and virtual dissection was performed in the whole hemisphere scan. Tract definitions can be found in a previous study (Maffei et al., 2021). We transferred these tract definitions to the block after registering the T2-weighted image from the MSME scan to the b=0 image from the whole hemisphere manually. The identified tracts in the tissue blocks were the superior longitudinal fasciculus (SLF), the anterior commissure (AC), the uncinate fasciculus (UF), and the internal capsule fibers projecting to superior frontal, inferior frontal, and anterior prefrontal cortex (IC1-3). We generated tract ROIs in the block space by applying the inverse affine matrix from the manual registration to the tracts and truncating them. The UF segment in the block was near the intersection with the extreme capsule, which is sometimes defined as the inferior fronto-occipital fasciculus (IFOF).

In the macaque brains, we analyzed the variation and correlations of microstructural features across the corpus callosum CC. We extracted CC ROIs by using the CVIM high-resolution ex vivo macaque brain diffusion MRI atlas (Calabrese et al., 2015). We registered the b=0 DW image from the atlas to the bias-corrected T2-weighted image in the MSME scan, and then applied the warp field to atlas tracts. We used the seven segments of the CC following the Witelson classification. These included the genu (label 'ccg'), segment I (label 'cc1d'), segment II (label 'cc2d'), segment III (label 'cc3d'), segment IV (label 'cc4d'), segment V (label 'cc5d'), and the splenium ('ccs').

## 3 Results

### 3.1 Quality of axon diameter model fit

We found a negligible effect of the denoising method on the powder-averaged signal decay curves and therefore on axon diameter index estimation. The evaluation can be found in supplementary materials (S4). Thus, in the following we report parameter estimates from the standard pipeline of denoising magnitude images.

Figure 1 shows the spherical mean signal decay versus b-value and the distribution of MCMC samples from a WM voxel in (a) the human block and (b) macaque sample 1. The axon diameter model fitted data from both species well (Fig. 1a and 1b). From the estimated parameters, we found that the human sample had a higher dot signal fraction (~0.3) than the macaque samples (~0.1), possibly due to differences in tissue preparation, e.g. fixation by immersion vs. perfusion. The parallel diffusivity $D_{\parallel}^{ia}$ is a tissue factor that affects the resolution limit for axon diameter index estimation (Andersson et al.,

2022; Drobnjak et al., 2016; Nilsson et al., 2017). The estimated $D_\parallel^{ia}$ was mostly around 0.4-0.5 $\mu m^2$/ms for the macaque samples and 0.6-0.75 $\mu m^2$/ms for the human tissue. Considering the different SNRs and estimated intrinsic diffusivities, we calculated the sensitivity ranges for the direction-averaged signals from a single b-value measurement of high gradient strength and n = 32, as introduced in (Andersson et al., 2022) and detailed in (Gong and Yendiki, 2024); the sensitivity range of our measurements to axon diameter index generally covers the 2 – 8 $\mu m$ range considering the multiple high b-values (Gong and Yendiki, 2024).

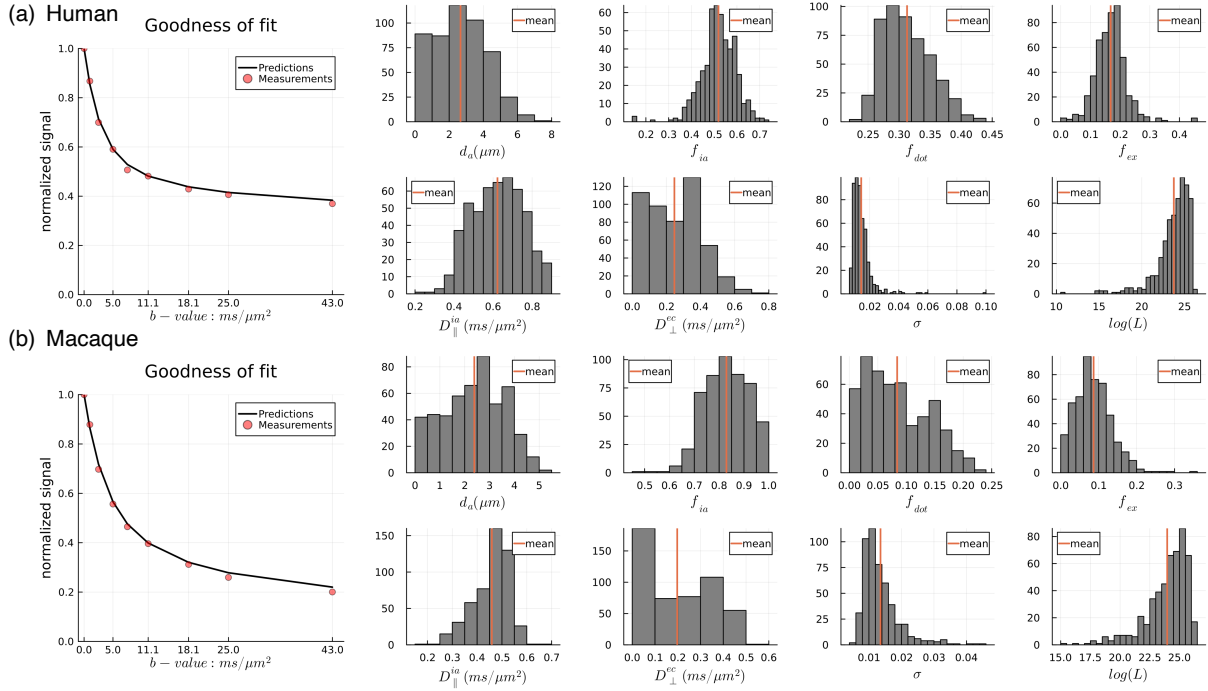

**Figure 1.** Quality of fitting for axon diameter index estimation. The spherical mean signal decay versus b-value and distribution of MCMC samples are shown from a WM voxel of the ALIC for the (a) human tissue block and (b) macaque brain 1. The human sample has a higher dot signal fraction.

Figure 2 shows fitting evaluations on synthetic data generated by the forward model in Eq. [1], to assess the accuracy of the axon diameter index estimate for different levels of SNR and ground-truth values of this model parameter. We used ground-truth axon diameter indices from 1 $\mu$m to 10 $\mu$m at 1 $\mu$m intervals. This range extends slightly beyond the sensitivity limits and is broadly representative axon diameter indices. The other tissue parameters chosen for the synthetic data were based on the estimates seen in (a) macaque: $f_{ia}$=0.8, $f_{dot}$=0.1, $D_\parallel^{ia}$ = 0.45 $\mu m^2$/ms and $D_\perp^{ec} = D_\parallel^{ia}*0.4$ and (b) human: $f_{ia}$=0.6, $f_{dot}$=0.3, $D_\parallel^{ia}$ = 0.65 $\mu m^2$/ms and $D_\perp^{ec} = D_\parallel^{ia}*0.4$. We added Gaussian noise to simulated signals in each case to generate 100 realizations with SNR levels of 150, 100 and 50 for signals synthesized with the macaque tissue properties and SNR of 150, 100 and 75 for signals synthesized with the human tissue properties. In all simulations, different axon diameter indices could be resolved, although with different levels of bias in the estimates. Decreased SNR levels introduced higher bias to estimates of axon diameter index and intra-axonal signal fractions. Increased dot signal fraction in the human tissue simulation decreased differentiation between smaller axon diameter indices. We use these simulation results in the following to help interpret estimates from real ex vivo data.

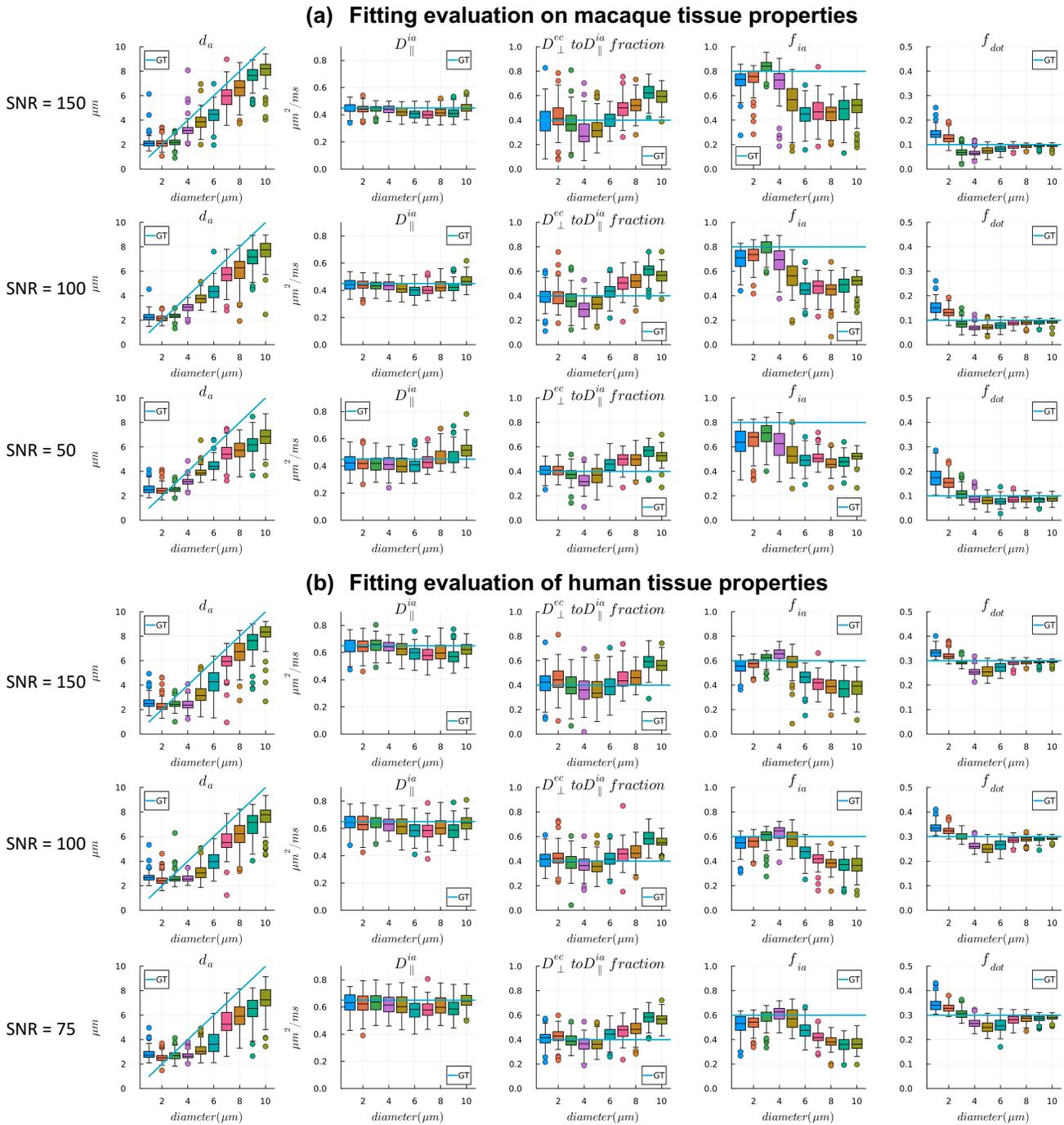

**Figure 2.** Parameter estimates from the axon diameter model, for data simulated with dot signal fraction and intrinsic diffusivities comparable to those from our ex vivo data in (a) macaque brains and (b) human sample. Three SNR levels were simulated in each case, to cover the range encountered in our data. Parameter estimates from 100 noise realizations are shown as boxplots and ground-truth (GT) parameter values are shown as line plots.

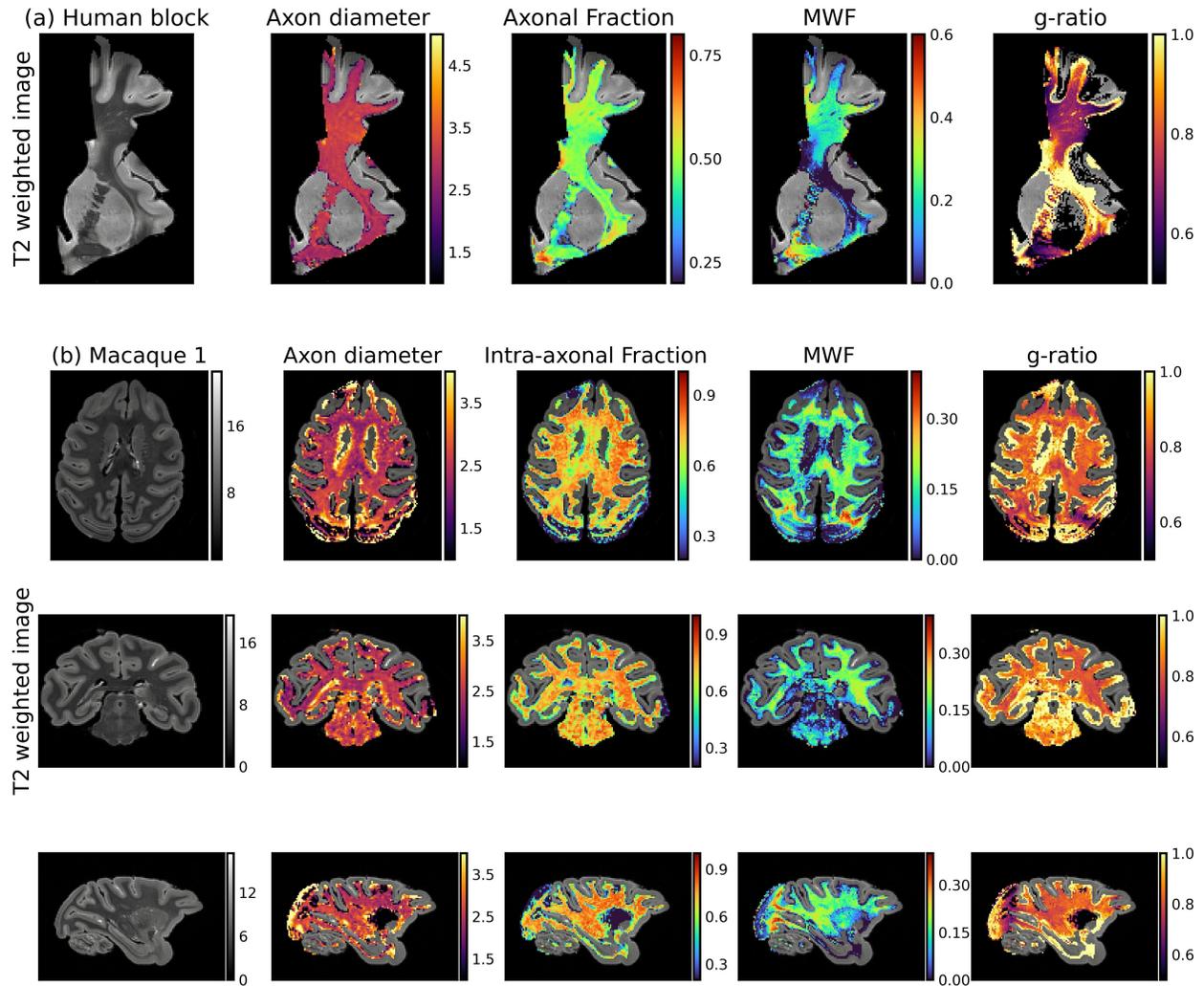

**Figure 3.** Microstructural parameter maps from the human block (a) and macaque sample 1 (b).

### 3.2 Variations of microstructure measures
Figure 3 shows example maps of the four microstructural parameters in the human block and macaque sample, which were quantitatively analyzed below across WM tracts and samples.

**Human sample**
Figure 4 shows the location of the TOIs from the human tissue slab and the distributions of estimated microstructural parameters within each TOI. The distributions of axon diameter indices showed small variations. Among the SLF, UF and AC, the SLF had the largest estimates of axon diameter index. The distributions of diameter index from segments of the IC were largely overlapping, with a small trend of larger diameter indices in fibers projecting to the superior frontal and inferior frontal (IC1-2) than the anterior prefrontal cortex (IC3). In all TOIs, the estimates of axon diameter index were mostly between

2.5-3.5 $\mu$m, and the intra-axonal tissue fractions were between 0.4-0.7. Compared to axon diameter index, the MWF had higher variability among TOIs, resulting in variability of the corresponding g-ratio. Some voxels had very low MWF as shown in Figure 3 (<0.05), e.g. the UF and IC, which led to peaks in the distributions of g-ratio near 1 while the majorities of the voxels had g-ratio between 0.6-0.8.

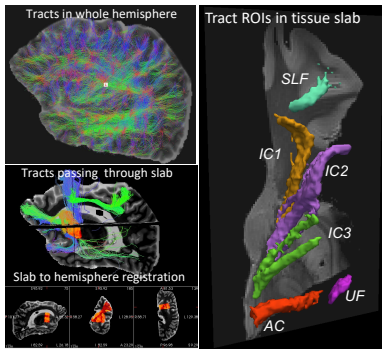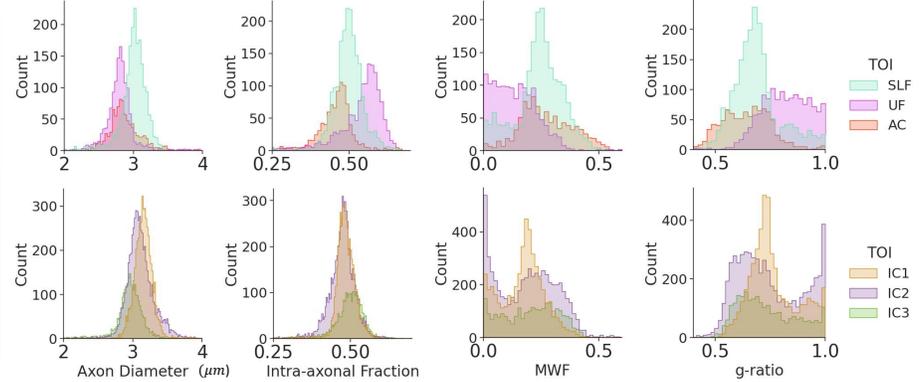

**Figure 4.** Distributions of parameters in TOIs of the human sample. (a) Pipeline for TOI identification: the whole left hemisphere scan was used for tractography and localization of the tissue block; the tissue block was registered to the whole hemisphere manually and TOIs passing through the block were identified and mapped to the block space. (b) Distributions of microstructural parameters in the identified TOIs. SLF = superior longitudinal fasciculus, AC = anterior commissure, UF = uncinate fasciculus, and IC1-3 = internal capsule fibers projecting to superior frontal, inferior frontal, and anterior prefrontal cortex.

**Macaque samples**

Figure 5 shows the distributions of microstructural parameters across CC segments as violin plots. We did not observe major variation of axon diameter index (mostly between 2-3 $\mu$m) or intra-axonal signal fractions (around 0.8) across segments of the CC, for any of the macaque samples. For MWF, the splenium of the CC had higher myelin concentrations than the body of the CC. All the ROIs were myelinated with MWF mostly between 0.1-0.3. This resulted in a reverse trend in g-ratio, which was mostly higher than 0.8.

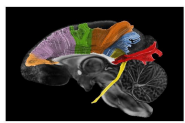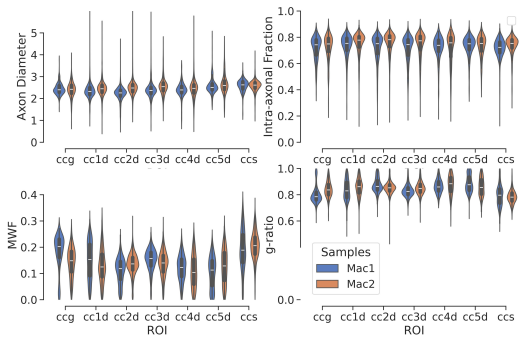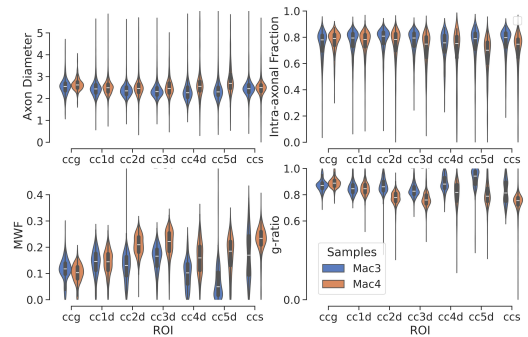

**Figure 5.** Distributions of microstructure parameters in the segments of the CC in the macaque brains; the unit for axon diameter index is $\mu m$.

## 3.3 Correlations between microstructural measures

Figure 6 shows the correlations between microstructural parameters, across voxels of all tracts shown in Figures 4 (frontal WM tracts in the human sample) and 5 (CC segments in the macaque brains). We excluded any voxels with nearly no myelination (MWF<0.05). We found consistent, albeit low correlations between axon diameter index and all other measures in the four higher SNR cases (WM SNR>30) except for the low SNR macaque 1. Specifically, axon diameter index was significantly negatively correlated with intra-axonal fraction and MWF across samples. Axon diameter index was positively correlated with g-ratio in the high SNR data, while this correlation is not consistently significant. The correlations between intra-axonal fraction and MWF were positive in the human sample but not consistently so in the macaque brains; the MWF had a wider range of values in the human sample. The g-ratio, while determined by MWF and axonal fraction, was mostly negatively correlated with MWF. Figure 7 further shows the density plot of estimates from the TOIs in each sample. We found consistent patterns between axon diameter index and the other measures in the higher SNR macaque and human samples.

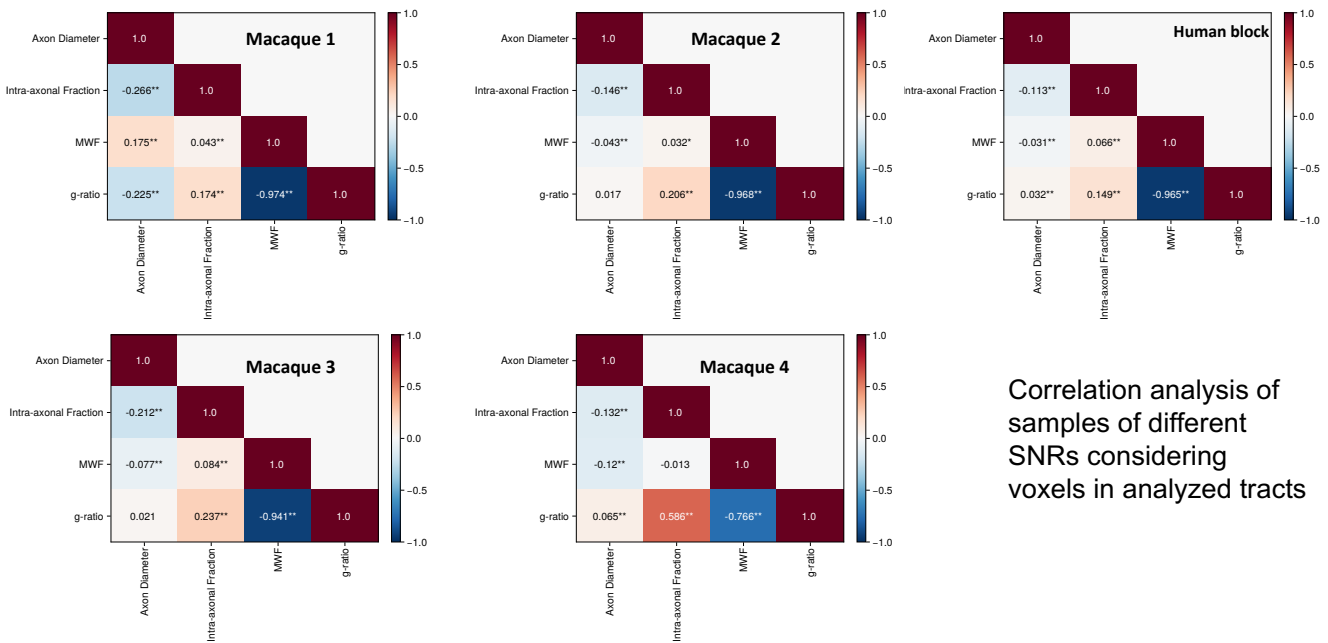

Correlation analysis of samples of different SNRs considering voxels in analyzed tracts

**Figure 6.** Parameter correlations in all the samples. The seven CC TOIs in the macaque samples and the six TOIs in the human block were included for analysis. Significant correlations were marked with ** ($p<0.01$) and * ($p<0.05$). Other than macaque sample 1, which has the lowest SNR, correlations between axon diameter index and other measures in all other samples were consistent.

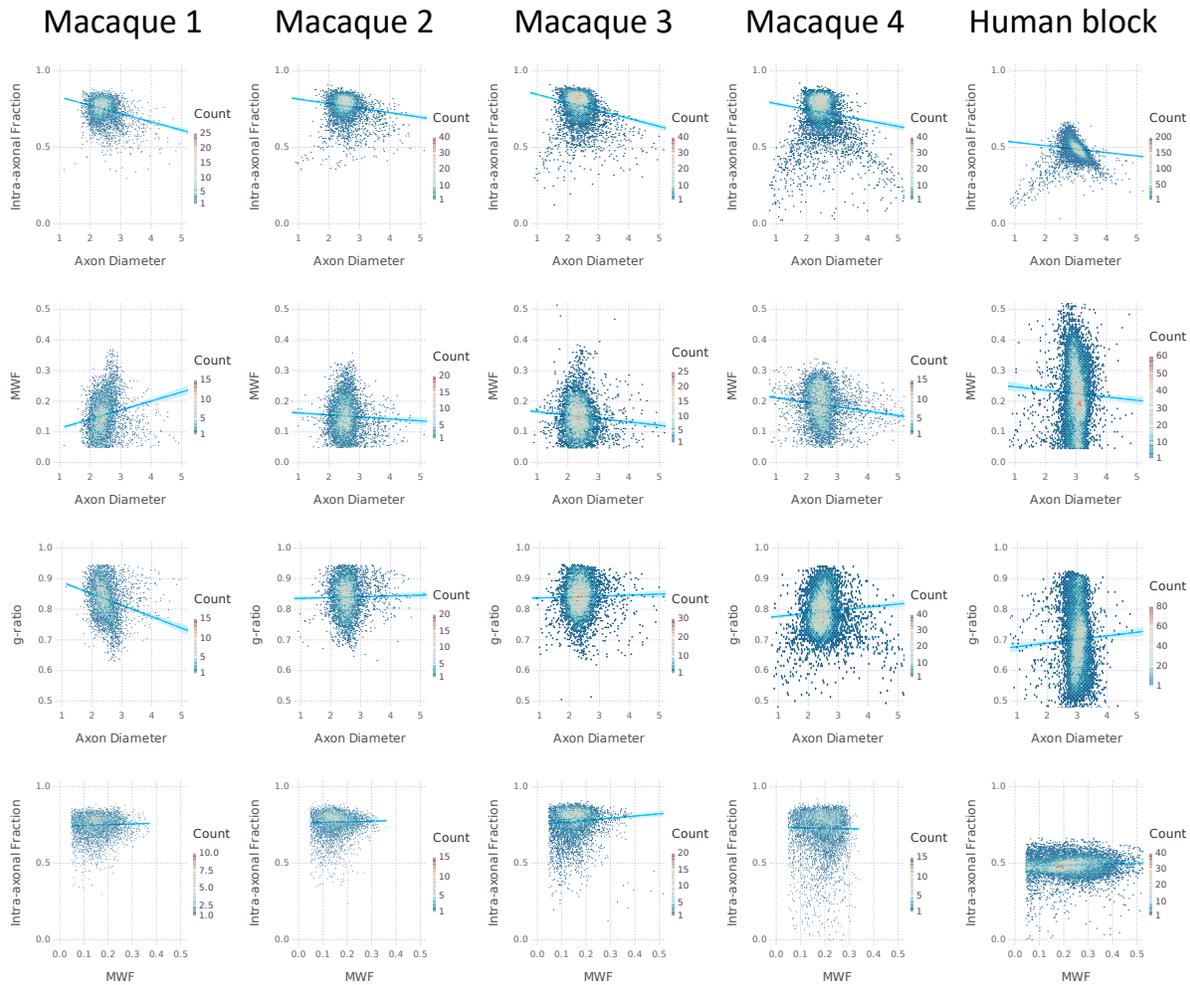

**Figure 7.** Density plots of estimates from all the voxels in TOIs. The seven CC TOIs in the macaque samples and the six TOIs in the human block were included. Other than macaque sample 1, which has the lowest SNR, correlations between axon diameter index and other measures in all other samples were consistent.

## 4 Discussion

In this study, we investigated the correlations between MRI-based estimates of axon diameter index and myelination in WM tracts across human and macaque brain samples. We estimated axon diameter index using a multi-compartment model with the spherical mean technique and evaluated fitting accuracy with synthetic data. We quantified myelin content using MWF and derived a measure of aggregate g-ratio by combining measures derived from dMRI and myelin imaging. We found consistent correlations between axon diameter index with other microstructural measures in samples with sufficient SNR. We discuss the biological implications and methodological considerations below.

### 4.1 Relationship between axon diameter index and myelination

Axon diameter index was weakly negatively correlated with intra-axonal signal fraction and MWF across tracts in samples with sufficient SNR, such that smaller axon estimates were associated with higher intra-axonal signal fraction and myelin concentration and a tendency of lower g-ratio. These findings are in accordance with existing evidence from electron microscopy studies. First, axons with diameter above 0.2 µm are myelinated in the healthy central nervous system (Hirano and Llena, 1995; Waxman and Bennett, 1972). Axons with smaller diameters are often associated with higher axonal packing density. For example, axons in the genu and splenium of CC appear smaller and denser than in the body of CC in both human and macaque (Aboitiz et al., 1992; Lamantia and Rakic, 1990). We acknowledge that the correlation between axon diameter and intra-axonal fraction may be an artifact of multi-compartment model fitting, and an almost linear negative correlation between axon diameter and intra-axonal signal fraction was also reported in other study (De Santis et al., 2016). Second, myelin thickness increases with axon diameter, leading to a relatively constant or increasing g-ratio (Hildebrand and Hahn, 1978; Keyserlingk and Schramm, 1984), which agrees with our finding. Given this constant/increasing g-ratio, tissue voxels with small and densely packed axons will have higher myelin concentration. In our MRI findings, we see the same trends, with slightly increasing trend of g-ratios when axon diameter indices increase. Furthermore, the axon diameter index and myelination estimates originate from imaging datasets of different contrasts (DW vs. T2-weighted), rendering their correlations more likely to be biologically meaningful rather than an artifact of model fitting. The correlations between axon diameter index and MWF could also have implications to studies investigating their individual contribution to quantitative relaxometry measures such as R1 (Harkins et al., 2016).

Surprisingly, while we found correlations between axon diameter index and myelination, we did not observe consistent correlations between intra-axonal signal fraction and MWF within well-myelinated CC tracts. While we acknowledge that estimates of intra-axonal signal fraction and MWF might suffer from different biases, an explanation for this lack of correlation is that our intra-axonal signal fraction could include contributions of restricted diffusion signals from other unmyelinated cylindrical structures, like unmyelinated axons and glia processes, albeit low in the WM. Although the myelin sheath has been considered the main barrier restricting water diffusion perpendicular to axons, early NMR studies also suggest that, even in the absence of myelin, cell membranes alone can introduce diffusion anisotropy (Beaulieu, 2002; Beaulieu and Allen, 1994). Direct evidence is provided by a study using very high-resolution imaging (9 µm) of excised lamprey spinal cord with large and unmyelinated axons (>20 µm), which shows that the diffusion signal is isotropic when measured entirely within a giant axon and anisotropic when measured near axon boundaries (Takahashi et al., 2002). This validates that axon membranes alone cause intra-axonal restriction. Previous studies that demonstrate a positive correlation between intra-axonal signal fraction and MWF (Billiet et al., 2015; De Santis et al., 2014), with exceptions (Qian et al., 2020), were performed under clinical acquisition conditions with much longer diffusion time and lower b-values, where the faster exchange between water in unmyelinated structures and extra-cellular space could have made them indistinguishable. The much shorter diffusion time and ultra-high b-values used in our experiments allow us to gain sensitivity to unmyelinated structures, as it has been demonstrated to separate restricted signal fractions between cell soma and neurites in the gray matter (Ianus et al., 2021; Palombo et al., 2020). However, diseases that affect membrane permeability might change this condition. Thus, combining diffusion measures with MWF is pertinent in the study of brain disease. Overall, this finding suggests that axonal geometry and MWF can provide complementary information about tissue morphology. This complementary information will be especially helpful in characterizing tissue changes that are caused by coexisting processes, such as axonal growth vs. myelination in healthy brain development, and axonal damage vs. demyelination in diseases such as multiple sclerosis (Yu et al., 2019).

## 4.2 Limitations and future work
**Estimation of axon diameter index**

Several limitations of estimating axon diameter index from MRI could have contributed to our findings. First, our study is limited by the fact that the axon diameter index represents only a summary metric derived from the full distribution of axon diameters in a tissue voxel. Specifically, this index is heavily weighted by the tail of the distribution, i.e., larger axons contribute more to the estimate (Burcaw et al., 2015; Neuman, 1974; Sepehrband et al., 2016; Veraart et al., 2020). This could explain the gap between MR-estimated diameter index (>2 $\mu m$) and histological measurements, where most axons are below 1 $\mu m$ with a positive skew in the distributions due to the presence of larger axons in a tissue voxel. While the absolute values of this index may deviate from histological measurements, its differences among WM regions have been shown to correlate with histological findings (Mordhorst et al., 2025; Veraart et al., 2021). The relationship between the axon diameter index and the full axon diameter distribution depends on the pulse width relative to the axon diameter and tissue intrinsic diffusivity (Appendix II), thus the strength of the tail-weighting can vary (Burcaw et al., 2015; Veraart et al., 2020). Although estimating a distribution of axon diameters has been previously demonstrated by assuming a parametric form of the distribution (Assaf et al., 2008), it adds more parameters to the model and further increases the complexity of an already complex optimization problem.

Second, the sensitivity range for axon diameter index estimation, and particularly the lower bound, i.e., the minimal axon diameter that can be resolved (Andersson et al., 2022; Drobnjak et al., 2016; Gong and Yendiki, 2024; Nilsson et al., 2017), is highly dependent on the acquisition protocol. Andersson et al. 2022 conducted a comprehensive simulation study with ex vivo and preclinical settings using both the same IAS model as in our study and the power law method from Veraart et al., 2020. A major finding was that axon diameters between 2-8 microns can be robustly estimated, while axon diameter estimates beyond this range will be highly biased for both methods. For assessing performance with larger axons, Andersson et al. 2022 segmented axons in the monkey corpus callosum and complex crossing fiber regions from large field of view X-ray nano-holotomography and performed fitting evaluations with Monte Carlo simulation in the segmented axonal substrates. Results suggest that axon diameters can be estimated through the IAS model in ex vivo setting with strong gradients despite fiber crossing. The segmented axons were mostly from the upper tail of the axon distributions (2-4 microns in the splenium and 3-7 microns in crossing fiber), as smaller axons could not be segmented with X-ray microscopy. As most axons in human and macaque tissue have diameters below the lower bound (Liewald et al., 2014), estimates in voxels with mostly these small axons could be biased and confound the relationships that we are investigating here. Nevertheless, our experimental conditions, including the lower diffusivity in ex vivo tissue and strong gradients of preclinical scanners, lead to a substantially reduced lower bound for diameter index estimation compared to typical in vivo settings. While lower diffusivity and higher gradients can also reduce sensitivity to larger axons (Andersson et al., 2022; Gong and Yendiki, 2024), the upper bound of the sensitivity range in our experiments still covers the expected range of axon diameter index in macaque and human tissue.

Finally, biases originating from multi-compartment model fitting could contribute to our findings. The plausibility of using measurements with one or a few ultra-high b-value (>=20 ms/$\mu$m2 for ex vivo tissue) to recover axon diameter index across the brain has been validated with realistic simulations (Andersson et al., 2022; Lee et al., 2024, 2020a) and histology (Veraart et al., 2020). These studies focus on the intra-axonal compartment. In our ex vivo data, the non-zero signal floor at high b-values suggests a non-negligible contribution from a dot compartment (Supplementary Materials S5) that varies spatially in our samples and thus should be modelled. We therefore opted to use a multi-compartment modelling framework to model the full signal decay, where we also include lower b-values to model a potential extra-cellular compartment. We performed a more detailed theoretical sensitivity analysis with our experimental protocols and tissue properties, and quality-of-fit evaluations for the multi-compartment model (Gong and Yendiki, 2024). Our results agree with findings in Andersson et al., 2022, showing that diameter indices in the 2–8-micron range can be estimated and discriminated from each other, although different levels of biases are present. The current model does not consider the T2 differences between tissue compartments, which could also lead to different levels of bias in the T2-weighted intra-axonal signal fractions in different samples (Gong et al., 2020; Lampinen et al., 2019; Veraart et al., 2017), given that T2 measured in the macaque and human tissue were quite different.

Based on our recent evaluation (Gong and Yendiki, 2024), acquiring additional measurements along the orthogonal dimension of diffusion time would improve the estimation precision of axon diameter index and intra-axonal signal fraction. With the current tissue model, however, the additional diffusion times should be short to reduce contributing factors that have been observed in longer diffusion time studies, such as time-dependent diffusion from extra-cellular space (Burcaw et al., 2015; De Santis et al., 2016; Lee et al., 2018), axonal undulation (Brabec et al., 2019; Lee et al., 2024, 2020a) and caliber variation (Lee et al., 2024, 2020a, 2020b). This also requires that the SNR of the data is sufficient to differentiate measurements between different diffusion times in the short diffusion time range. Satisfying these requirements would lead to much prolonged data acquisition time. Considering that shorter diffusion times will also allow shorter echo times and hence improve SNR, more work is needed to identify the acquisition and analysis strategies that maximize estimation accuracy and precision. In the future, we plan to further investigate the optimal number of diffusion times, b-values, and number of gradient directions in a multi-dimensional acquisition with feasible acquisition time, as well as analysis strategies such as compensating for spatially correlated noise (Henriques et al., 2023) and improving spherical mean signal estimation (Afzali et al., 2021).

**Estimation of myelin concentration**
We have chosen to use myelin water imaging to quantify myelination, as MWF has been shown to be a highly sensitive myelination measure in the WM, when compared to several other methods (Does, 2018; Faizy et al., 2020; Sandrone et al., 2023). We use a relatively mature approach to estimate MWF by acquiring multiple spin-echo images and fitting a T2 spectrum. Several factors will affect the accuracy of the MWF estimation. First, water exchange between myelin water and intra/extra-axonal compartments has been shown to cause changes in the T2 spectrum and therefore lead to underestimating MWF (Harkins et al., 2012). This underestimation is more likely in smaller axons with faster water exchange due to their higher surface-to-volume ratio and thinner myelin sheath (Dortch et al., 2013). However, as we find that smaller axons are associated with higher MWF, water-exchange likely has negligible contribution to our findings. Second, the T2 spectrum fitting is very sensitive to SNR (MacKay et al., 2006); variations in SNR may contribute to the variations of MWF estimates in this study. For future studies to eliminate this fitting bias, a promising technique is the direct visualization of the short-T2 signal component by suppressing signals from other tissue components in the data acquisition (Oh et al., 2013). This technique has recently been accelerated by combination with MR fingerprinting (Liao et al., 2023). Finally, MWF estimates have been shown to be fiber orientation dependent (Birkl et al., 2021). This bias comes from orientation dependent T2, explained by a susceptibility model in myelinated adult brain (Wharton and Bowtell, 2012) and a residual dipolar coupling model in almost unmyelinated newborn brain (Bartels et al., 2022). Calibrating this orientation-dependence bias will be promising in myelin imaging, as has been recently demonstrated in diffusion-informed myelin water imaging (Chan and Marques, 2020) and quantitative susceptibility imaging (Sandgaard et al., 2024).

**Ex vivo tissue**

In comparison to in vivo tissue, ex vivo tissue provides several advantages for observing the relationship between microstructure measures. The lower diffusivity in ex vivo tissue allows us to approach lower resolution limits (Andersson et al., 2022; Drobnjak et al., 2016; Gong and Yendiki, 2024; Nilsson et al., 2017), thus gain sensitivity to smaller axons. Fixation, shown to increase intracellular water residence time (Thelwall et al., 2006), could lower the rate of water exchange, therefore making modelling assumptions, for both axon diameter index and myelin imaging, more reasonable. Nevertheless, fixation and related tissue preparation processes might also change tissue microstructure properties depending on the quality of tissue preservation.

Different fixation methods in macaque vs. human could contribute to the differences observed between species, i.e. the lower intra-axonal signal fraction and higher dot signal fraction in the human tissue. Different fixation methods may alter the physical and MR properties of tissue differently, such as membrane permeability, water diffusivity and T2 (Dadar et al., 2024; D'Arceuil et al., 2007; Pfefferbaum et al., 2004; Yendiki et al., 2022), and therefore differences in the observed compartment signal fractions that will be affected by these tissue properties. A previous electron microscopic study has reported a higher axon packing density in macaque than human tissue (Liewald et al., 2014). The explanation for this is the higher fixation quality for macaque brains, which were perfusion fixed, when compared to the immersion fixation performed on human brains. This study further explains that lower packing density does not imply a great loss of fiber populations but is partially due to disintegration of cellular material, evidenced by the debris in the space between myelinated fibers in human tissue and an overall similar distribution and mean values of axon diameter in both species. We similarly found lower intra-axonal signal fractions in the human tissue but comparable axon diameter indices in macaque and human brains. Another difference that we found is the higher variability in MWF estimates in the human tissue, which could also partly be due to differences in tissue preparations.

**5 Conclusion**

In this study, we used a preclinical system with $G_{max}$=660 mT/m to investigate relationships between MRI-derived estimates of microstructure measures with ex vivo brain samples in macaque and human tissue. The weak correlations between dMRI metrics and MWF suggest that they provide complementary information about fiber geometry and thus should be combined for thorough microstructure analysis. Validating our preliminary findings in vivo will be crucial to further our understanding of these effects. With the development of human MRI scanners with ultra-strong gradients, such as the Connectome 2.0 with $G_{max}$=500 mT/m (Huang et al., 2021), we will be able to assess axon diameter index and myelination across the brain with higher sensitivity compared to previous in vivo studies. Considering these correlations in healthy tissue as a baseline will be helpful when studying microstructural changes in disease. Another open question is whether microstructure measures can help differentiate fiber tracts as has been suggested to inform dMRI tractography (Battocchio et al., 2022; Girard et al., 2017; Schiavi et al., 2022). Future work will explore what acquisition protocols and microstructure measures are better suited for this purpose.

**Appendix I**

The powder-averaging of DW signals over gradient directions in the spherical mean technique (SMT) yields (Callaghan et al., 1979):

$$S_{smt} = e^{-bD_\perp} \sqrt{\frac{\pi}{4b(D_\parallel - D_\perp)}} \, \text{erf}\left(\sqrt{b(D_\parallel - D_\perp)}\right). \quad [A1]$$

For the intra-axonal compartment, the perpendicular diffusivity can be used to calculate the axon diameter index through Gaussian phase distribution approximation (Neuman, 1974). Therefore, the intra-axonal signal decay is (Fan et al., 2020; Jespersen et al., 2007; Kaden et al., 2016; Van Gelderen et al., 1994):

$$S_{ia}(b(\delta, \Delta, G); D_\parallel^{ia}, D_0, d_a) = e^{-bD_\perp^{ia}} \sqrt{\frac{\pi}{4b(D_\parallel^{ia} - D_\perp^{ia})}} \, \text{erf}\left(\sqrt{bD_\parallel^{ia} - bD_\perp^{ia}}\right);$$

$$-bD_\perp^{ia} = -2\gamma^2 G^2 \sum_{m=1}^{\infty} \frac{1}{D_0^2 \alpha_m^6 (r_a^2 \alpha_m^2 - 1)} \left(2D_0 \alpha_m^2 \delta - 2 + 2e^{-D_0 \alpha_m^2 \delta} + 2e^{-D_0 \alpha_m^2 \Delta} - e^{-D_0 \alpha_m^2 (\Delta - \delta)} - e^{-D_0 \alpha_m^2 (\Delta + \delta)}\right), \quad [A2]$$

where $r_a = d_a/2$ is the radius, $\alpha_m$ is the $m$-th root of $J_1'(\alpha r_a)=0$ and $J_1'$ is the derivative of the first-order Bessel function of the first kind. We calculated the contribution up to $m$=10 in this study. We use the full functional form instead of Neuman's limit for converting the perpendicular diffusivity to axon diameter index considering that it offers applicability to a wider range of axon diameters and intrinsic diffusivities (Andersson et al., 2022).

The anisotropic extra-cellular compartment takes the same form of spherical mean as:

$$S_{ec}(b; D_\perp^{ec}, D_\parallel^{ec}) = e^{-bD_\perp^{ec}} \sqrt{\frac{\pi}{4b(D_\parallel^{ec} - D_\perp^{ec})}} \, \text{erf}\left(\sqrt{b(D_\parallel^{ec} - D_\perp^{ec})}\right). \quad [A3]$$

**Appendix II**

The axon diameter index $d_a$ is a representative measure from a distribution of axon diameters within the tissue voxel. Given axons with a radius distribution of h(r), the definition of $d_a$ depends on the relation between pulse width ($\delta$) and the time required for a spin to diffuse across an axon ($r^2/D_0$). At the wide pulse limit, $\delta \gg r^2/D_0$, it depends on the ratio between the sixth and second moment of the distribution h(r) (Veraart et al., 2020):

$d_a = 2(\frac{\langle r^6 \rangle}{\langle r^2 \rangle})^{1/4}$ . [A4]

At the narrow pulse limit $\delta \ll r^2/D_0$, it depends on the ratio between the fourth and second moment of h(r) (Burcaw et al., 2015):

$d_a = 2(\frac{\langle r^4 \rangle}{\langle r^2 \rangle})^{1/2}$. [A5]

Hence, $d_a$ is heavily weighted by larger axons (the tail of the distribution) but is less tail-weighted and hence a somewhat better approximation for the mean of the distribution at the narrow pulse limit. Equations [A4-5] apply to both a distribution of diameters from axons represented by cylinders and a distribution of diameters from a single axon with a varying diameter along its trajectory (Lee et al., 2020a).